\documentclass[12pt]{article}
\usepackage[latin1]{inputenc}
\begin{document}

\baselineskip 0.85cm
\topmargin -0.55in
\oddsidemargin -0.2in

\let\ni=\noindent

\renewcommand{\thefootnote}{\fnsymbol{footnote}}

\newcommand{ \mi }{ {\stackrel{o}{m}}}

\newcommand{\SM}{Standard Model }

\pagestyle {plain}

\setcounter{page}{1}

\pagestyle{empty}

~~~

\begin{flushright}
IFT/00--12
\end{flushright}

\vspace{0.3cm}

{\large\centerline{\bf Two sterile neutrinos as a consequence}}

\vspace{0.1cm}

{\large\centerline{\bf of matter structure{\footnote{ Work supported in part by the
Polish KBN--Grant 2 P03B 052 16 (1999--2000).}}}}

\vspace{0.8cm}

{\centerline {\sc Wojciech Kr\'{o}likowski}}

\vspace{0.8cm}

{\centerline {\it Institute of Theoretical Physics, Warsaw University}}

{\centerline {\it Ho\.{z}a 69,~~PL--00--681 Warszawa, ~Poland}}

\vspace{1.0cm}

{\centerline{\bf Abstract}}

\vspace{0.3cm}

  An algebraic argument based on a series of generalized Dirac equations, 
truncated by an "intrinsic Pauli principle", shows that there should exist two 
sterile neutrinos as well as three families of leptons and quarks. Then, the 
influence of these additional neutrinos on neutrino oscillations is studied. 
As an example, a specific model of effective five--neutrino texture is 
proposed, where only the nearest neighbours in the sequence of five neutrinos 
ordered as $ \nu_s\,,\, \nu'_s\,,\, \nu_e\,,\, \nu_\mu\,,\, \nu_\tau $ are 
coupled through the $ 5\times 5 $ mass matrix. Its diagonal elements are 
taken as negligible in comparison with its nonzero off--diagonal entries.

\vspace{0.3cm}

\ni PACS numbers: 12.15.Ff , 14.60.Pq , 12.15.Hh .

\vspace{2.3cm}

\ni April 2000

\vfill\eject

~~~
\pagestyle{plain}

\setcounter{page}{1}

\vskip0.3cm

\ni {\bf 1. Introduction}

\vspace{0.3cm}

 The   existence problem of    light  sterile neutrinos  [1], free  of
 Standard Model gauge  charges,  is connected phenomenologically  with
 the LSND  effect for  accelerator  $\nu_\mu$'s [2].  If confirmed, it
 would  avail (jointly with the observed  deficits  of solar $\nu_e$'s
 [3] and  atmospheric $\nu_\mu$'s [4])    the existence of  the  third
 mass--square    difference    in     neutrino oscillations,  invoking
 necessarily at least one kind of light sterile neutrinos, in addition
 to   the          familiar     three        active          neutrinos
 $\nu_e\,,\,\nu_\mu\,,\,\nu_\tau  $.  From the theoretical  viewpoint,
 however, {\it light} sterile neutrinos might  exist even in the case,
 when the LSND effect was not confirmed [5].  At any rate, there might
 be {\it either} three sorts of light Majorana sterile neutrinos 

\begin{equation}
\nu_\alpha^{(s)} = \nu_{\alpha R} +(\nu_{\alpha R} )^c \;\;\;
(\alpha = e\,,\,\mu\,,\,\tau )
\end{equation}

\ni being structurally righthanded counterparts of familiar light 
Majorana active neutrinos

\begin{equation}
\nu_\alpha^{(a)} = \nu_{\alpha L} +(\nu_{\alpha L} )^c \;\;\;
(\alpha = e\,,\,\mu\,,\,\tau )\,,
\end{equation}

\ni {\it or} some quite new Dirac or Majorana light sterile neutrinos.

The first kind  appears in   the case   of pseudo--Dirac option    for
 neutrinos  $\nu_e\,,\,\nu_\mu\,,\,\nu_\tau $  [6] that contrasts with
 the popular seesaw option [7] involving {\it heavy} sterile neutrinos
 of the form (1). If the LSND effect did not exist, the seesaw option,
 operating effectively at low energies with three active neutrinos (2)
 only, would be phenomenologically  most economical, beside the simple
 option of Dirac neutrinos $\nu_\alpha  = \nu_{\alpha L} + \nu_{\alpha
 R}$. At the same time, the seesaw option would be favourable from the
 standpoint of  GUT idea (say, in the  SO(10) version), where  a large
 mass scale of sterile neutrinos of  the form (1) could be understood,
 at least  qualitatively. On the other  hand, however,  such an option
 would not meet the needs of  astrophysics for light sterile neutrinos
 useful in tentatively explaining heavy--element nucleosynthesis [8]. 

 The sterile neutrinos  of the second kind are  suggested to exist [9]
 in the framework of  a theoretical scheme   based on the series $N  =
 1,2,3,\ldots$ of generalized Dirac equations [10] 

\begin{equation}
\left\{ \Gamma^{(N)}\cdot\left[p - g A(x)\right] - M^{(N)}\right\} 
\psi^{(N)}(x) = 0\,,
\end{equation}

\ni where for any $N$ the Dirac algebra

\begin{equation}
\left\{ \Gamma^{(N)}_\mu\,,\,\Gamma^{(N)}_\nu \right\} = 2 g_{\mu \nu}
\end{equation}

\ni is constructed by means of a Clifford algebra,

\begin{equation}
\Gamma^{(N)}_\mu \equiv \sum^N_{i=1}  \gamma^{(N)}_{i \mu}\;\; , \;\;
\left\{ \gamma^{(N)}_{i \mu}\,,\,\gamma^{(N)}_{j \nu} \right\} 
= 2 \delta_{i j} g_{\mu \nu}
\end{equation}

\ni with $i\,,\,j = 1,2,\ldots ,N$ and  $\mu\,,\,\nu = 0,1,2,3$. Here,
the term   $g \Gamma^{(N)} \cdot  A(x)$  symbolizes the Standard Model
gauge  coupling,  involving  $\Gamma^{(N)}_5    \equiv i\Gamma^{(N)}_0
\Gamma^{(N)}_1  \Gamma^{(N)}_2 \Gamma^{(N)}_3$ as  well  as the color,
weak--isospin and  hypercharge  matrices (this coupling is  absent for
sterile particles such  as sterile neutrinos). The  mass $ M^{(N)}$ is
independent of $\Gamma^{(N)}_\mu$.   In general, the  mass $  M^{(N)}$
should be  replaced by a mass matrix  of elements  $ M^{(N,N')}$ which
would couple $\psi^{(N)}(x)$  with all  appropriate  $\psi^{(N')}(x)$,
and   it might  be natural to   assume  for $N  \neq  N'$ that $\left[
\gamma^{(N)}_{i  \mu}\, , \, \gamma^{(N')}_{j  \nu}  \right] = 0$ {\it
i.e.}, $\left[ \Gamma^{(N)}_\mu\, , \, \Gamma^{(N')}_\nu \right] = 0$. 

 The Dirac--type equation (3) for any $N$ implies that

\begin{equation}
\psi^{(N)}(x) = 
\left( \psi_{\alpha_1\alpha_2 \ldots \alpha_N}^{(N)}(x) \right) \,,
\end{equation}

\ni where each $\alpha_i = 1,2,3,4 $ is the Dirac bispinor index defined in 
its chiral representation in which the matrices

\begin{equation}
\gamma^{(N)}_{j 5} \equiv i \gamma^{(N)}_{j 0} \gamma^{(N)}_{j 1} 
\gamma^{(N)}_{j 2} \gamma^{(N)}_{j 3} \;\; , \;\; \sigma^{(N)}_{j 3} 
\equiv \frac{i}{2} \left[ \gamma^{(N)}_{j 1} \, , \, 
\gamma^{(N)}_{j 2} \right]
\end{equation}

\ni  are diagonal  (note that all  matrices (7),  both  with equal and
different $j$'s, commute simultaneously).  The wave function  or field
$\psi^{(N)}(x)  $   for  any $N$   carries  also   the  Standard Model
(composite)  label,  suppressed in our  notation.  The mass $ M^{(N)}$
gets also such a label. The Standard Model  coupling of physical Higgs
bosons should be eventually added to Eq. (3) for any $N$. 

 For $ N = 1$ Eq. (3) is, of course, the usual Dirac equation, for $ N
 = 2$ it is  known as the Dirac  form [11] of the  K\"{a}hler equation
 [12], while  for $   N \geq 3$   Eqs.  (3) give  us  new  Dirac--type
 equations [10].  All   of them  describe   some spin--halfinteger  or
 spin--integer particles for $N$  odd and $N$ even,  respectively. The
 nature of  these particles is the  main subject  of the present paper
 ({\it cf.} also Ref. [10]). 

 The Dirac--type matrices $\Gamma^{(N)}_\mu $ for any $N$ can be 
embedded into the new Clifford algebra

\begin{equation}
\left\{ \Gamma^{(N)}_{i \mu}\, ,  \,\Gamma^{(N)}_{j \nu} \right\} 
= 2\delta_{i j} g_{\mu \nu}
\end{equation}

\ni [isomorphic  with    the    Clifford  algebra   introduced     for
$\gamma^{(N)}_{i  \mu}$  in Eq.  (5)],  if $\Gamma^{(N)}_{i  \mu}$ are
defined by the  properly  normalized  Jacobi linear  combinations   of
$\gamma^{(N)}_{i \mu}$. In fact, they are given as 

\begin{equation}
\Gamma^{(N)}_{1 \mu} \equiv \Gamma^{(N)}_\mu \equiv \frac{1}{\sqrt{N}} 
\sum^N_{i=1}  \gamma^{(N)}_{i \mu}\,\; ,\,\;\Gamma^{(N)}_{i \mu} 
\equiv \frac{1}{\sqrt{i(i-1)}} \left[ \gamma^{(N)}_{1 \mu} + \ldots + 
\gamma^{(N)}_{(i-1) \mu} - (i - 1) \gamma^{(N)}_{i \mu} \right] 
\end{equation}

\ni   for $  i   =  1$ and  $  i  =   2,\ldots, N$, respectively.  So,
$\Gamma^{(N)}_{1}$ and         $\Gamma^{(N)}_{2}      \,      ,     \,
\ldots,\Gamma^{(N)}_{N}$ represent respectively the "centre--of--mass"
and  "relative"  Dirac--type  matrices.   Note that  the   Dirac--type
equation (3)  for any $N$ does  not involve the "relative" Dirac--type
matrices  $\Gamma^{(N)}_{2}  \, ,  \, \ldots,\Gamma^{(N)}_{N}$, solely
including  the "centre--of--mass" Dirac--type matrix $\Gamma^{(N)}_{1}
\equiv \Gamma^{(N)}$. Since $\Gamma^{(N)}_{i}  = \sum^N_{j=1} O_{i  j}
\gamma^{(N)}_{j}$, where the  $N\times N$ matrix $  O = \left( O_{i j}
\right)$ is orthogonal ($O^T = O^{-1}$), we  obtain for the total spin
tensor the formula 

\begin{equation}
\sum^N_{i=1}  \sigma^{(N)}_{i \mu \nu} = \sum^N_{i=1}  
\Sigma^{(N)}_{i \mu \nu}  \,,
\end{equation}

\ni where

\begin{equation} 
\sigma^{(N)}_{j \mu \nu} \equiv \frac{i}{2} \left[ \gamma^{(N)}_{j \mu} 
\, , \, \gamma^{(N)}_{j \nu} \right] \;\; ,\;\; \Sigma^{(N)}_{j \mu \nu} 
\equiv \frac{i}{2} \left[ \Gamma^{(N)}_{j \mu} \, , \, 
\Gamma^{(N)}_{j \nu} \right]\,.
\end{equation}

\ni Of course, the spin tensor (10) is the generator of Lorentz 
transformations for $\psi^{(N)}(x)$.

 It is convenient for any $N$ to  pass from the chiral representations
 for individual  $\gamma^{(N)}_i$'s to the  chiral representations for
 Jacobi $\Gamma^{(N)}_i$'s in which the matrices 

\begin{equation}
\Gamma^{(N)}_{j 5} \equiv i\Gamma^{(N)}_{j 0} \Gamma^{(N)}_{j 1} 
\Gamma^{(N)}_{j 2} \Gamma^{(N)}_{j 3} \;\; ,\;\; \Sigma^{(N)}_{j 3} 
\equiv \frac{i}{2} \left[ \Gamma^{(N)}_{j 1} \, , \, \Gamma^{(N)}_{j 2}\right]
\end{equation}

\ni are diagonal  (they  all, both   with equal  and different  $j$'s,
commute    simultaneously). Note  that   $\Gamma^{(N)}_{1   5}  \equiv
\Gamma^{(N)}_5 $ is the Dirac--type chiral matrix as it is involved in
the Standard Model gauge coupling in the Dirac--type equation (3). 

 Using the  new Jacobi chiral representations,  the "centre--of--mass"
 Dirac-type  matrices  $\Gamma^{(N)}_{1 \mu}  \equiv \Gamma^{(N)}_\mu$
 and $ \Gamma^{(N)}_{1  5} \equiv \Gamma^{(N)}_5$ can  be taken in the
 reduced forms 

\begin{equation}
\Gamma^{(N)}_\mu =  \gamma_\mu \otimes \underbrace{ {\bf 1}\otimes \cdots 
\otimes {\bf 1}}_{ N-1 \;{\rm times}} \;\; , \;\; \Gamma^{(N)}_5 = \gamma_5  
\otimes \underbrace{ {\bf 1}\otimes \cdots \otimes 
{\bf 1}}_{ N-1 \;{\rm times}} \; , 
\end{equation}

\ni where $\gamma_\mu$, $ \gamma_5 \equiv i \gamma_0 \gamma_1 \gamma_2
\gamma_3 $ and {\bf  1} are the  usual $4\times 4$ Dirac matrices. For
instance,  the Jacobi   $\Gamma^{(N)}_{i \mu}$'s and  $\Gamma^{(N)}_{i
5}$'s for $N = 3$ can be chosen as 

$$
\begin{array}{lllllll}
\Gamma^{(3)}_{1 \mu} & = & \gamma_\mu \otimes {\bf 1}\otimes  {\bf 1} & , & 
\Gamma^{(3)}_{1 5} & = & \gamma_5 \otimes {\bf 1}\otimes  {\bf 1} \; , \\ \\
\Gamma^{(3)}_{2 \mu} & = & \gamma_5 \otimes i \gamma_5 \gamma_\mu \otimes  
{\bf 1} & , & 
\Gamma^{(3)}_{2 5} & = & {\bf 1} \otimes \gamma_5 \otimes {\bf 1} \; , \\ \\
\Gamma^{(3)}_{3 \mu} & = & \gamma_5 \otimes \gamma_5 \otimes \gamma_\mu & , & 
\Gamma^{(3)}_{3 5} & = &  {\bf 1}  \otimes {\bf 1}\otimes  \gamma_5 \; . 
\end{array}
$$

\vspace{-1.2cm}

\begin{flushright}
(14)
\end{flushright}

\addtocounter{equation}{+1}

\vspace{0.2cm}

 Then, the Dirac--type equation (3) for any $N$ can be rewritten in 
the reduced form

\begin{equation}
\left\{ \gamma \cdot \left[p - g A(x)\right] - 
M^{(N)}\right\}_{\alpha_1\beta_1} 
\psi^{(N)}_{\beta_1 \alpha_2 \ldots \alpha_N}(x) = 0\;,
\end{equation}

\ni where $\alpha_1$ and $\alpha_2 \,,\, \ldots\,,\, \alpha_N$ are the
"centre--of--mass" and "relative" Dirac bispinor indices, respectively
(here, $(\gamma \cdot p)_{\alpha_1\beta_1} =  \gamma_{\alpha_1\beta_1}
\cdot   p$     and  $\left(  M^{(N)}     \right)_{\alpha_1\beta_1}   =
\delta_{\alpha_1\beta_1} M^{(N)}$, but the  chiral coupling  $g \gamma
\cdot A(x)$ involves within $A(x)$ also the  matrix $\gamma_5$ ). Note
that  in the Dirac--type  equation (15) for  any  $N>1$ the "relative"
indices $\alpha_2 \,,\, \ldots\,,\, \alpha_N$ are  free, but still are
subjects of  Lorentz transformations  (for $\alpha_2$  this  was known
already in the  case of Dirac   form [11] of K\"{a}hler  equation [12]
corresponding to our $N = 2$). 

Since in Eq. (15) the  Standard Model gauge  fields interact only with
the "centre--of--mass"  index $\alpha_1$, this  is distinguished  from
the    physically  unobserved   "relative"    indices $\alpha_2  \,,\,
\ldots\,,\, \alpha_N$. Thus, it was  natural for us to conjecture some
time    ago  that  the "relative"  bispinor   indices  $\alpha_2 \,,\,
\ldots\,,\, \alpha_N$   are  all   undistinguishable  physical objects
obeying  Fermi statistics along  with the Pauli principle requiring in
turn the  full antisymmetry of  wave function $\psi_{\alpha_1 \alpha_2
\,,\,  \ldots\,,\,  \alpha_N}(x)$  with   respect to  $\alpha_2  \,,\,
\ldots\,,\, \alpha_N$ [10]. Hence, only  five values of $N$ satisfying
the condition $N-1\leq 4$ are allowed, namely  $N = 1,3,5$ for $N$ odd
and $N = 2,4$ for $N$ even. Then, from the postulate of relativity and
the probabilistic interpretation of  $\psi^{(N)}(x)$  we were able  to
infer that  three $N$ odd and  two $N$ even  correspond to states with
total spin 1/2 and total spin 0, respectively [10]. 

 Thus, the Dirac--type equation (3), jointly with the "intrinsic Pauli
 principle", if  considered  on  a  fundamental  level, justifies  the
 existence in   Nature  of {\it three  and    only three}  families of
 spin--1/2   fundamental fermions ({\it   i.e.},  leptons  and quarks)
 coupled to the \SM gauge bosons. In addition, there should exist {\it
 two and only two} families of spin--0 fundamental bosons also coupled
 to the \SM gauge bosons. 

 For sterile particles, Eq. (15) with any $N$ goes over into the free 
Dirac--type equation

\begin{equation}
\left(\gamma_{\alpha_1\beta_1} \cdot p - \delta_{\alpha_1\beta_1} 
M^{(N)}\right) \psi^{(N)}_{\beta_1 \alpha_2 \ldots \alpha_N}(x) = 0
\end{equation}

\ni (as far  as only \SM gauge  interactions are considered). Here, no
Dirac bispinor index $\alpha_i$   is  distinguished by the  \SM  gauge
coupling  which is absent in   this case. The "centre--of mass"  index
$\alpha_1$ is not distinguished also by its coupling to the particle's
four--momentum, since Eq.  (16)  is physically equivalent to  the free
Klein--Gordon equation 

\begin{equation}
\left( p^2 - M^{(N)\,2}\right) 
\psi^{(N)}_{\alpha_1 \alpha_2 \ldots \alpha_N}(x) = 0 \;.
\end{equation}

\ni Thus, in this case the intrinsic  Pauli principle requires that $N
\leq 4$, leading to $N = 1,3$ for $N$ odd and $N  = 2,4$ for $N$ even.
Similarly as before, they correspond to states with total spin 1/2 and
total spin 0, respectively [9]. 

 Therefore,  there should exist  {\it   two  and only two}   spin--1/2
 sterile  fundamental   fermions ({\it i.e.},  two  sterile  neutrinos
 $\nu_s$  and  $\nu'_s$)  and, in   addition, {\it  two  and only two}
 spin--0 sterile fundamental bosons. 

 The wave functions or fields of active  fermions (leptons and quarks)
of  three families and  sterile  neutrinos of two  generations  can be
presented  in terms      of  $\psi^{(N)}_{\alpha_1 \alpha_2     \ldots
\alpha_N}(x)$ as follows 

\begin{eqnarray} 
\psi^{(f)}_{\alpha_1}(x) & = & \psi^{(1)}_{\alpha_1}(x) \;, \nonumber \\
\psi^{(f')}_{\alpha_1}(x) & = & \frac{1}{4}\left(C^{-1} 
\gamma_5 \right)_ {\alpha_2 \alpha_3} 
\psi^{(3)}_{\alpha_1 \alpha_2 \alpha_3}(x) = 
\psi^{(3)}_{\alpha_1 1 2}(x) = \psi^{(3)}_{\alpha_1 3 4}(x) \;,\nonumber \\
\psi^{(f'')}_{\alpha_1}(x) 
& = & \frac{1}{24}\varepsilon_{\alpha_2 \alpha_3 \alpha_4 \alpha_5} 
\psi^{(5)}_{\alpha_1 \alpha_2 \alpha_3 \alpha_4 \alpha_5}(x) = 
\psi^{(5)}_{\alpha_1 1 2 3 4}(x) 
\end{eqnarray} 

\ni and

\begin{eqnarray} 
\!\!\! \psi^{(\nu_s)}_{\alpha_2}(x) & = & 
\psi^{(1)}_{\alpha_2}(x) \;, \nonumber \\
\!\!\! \psi^{(\nu'_s)}_{\alpha_2}(x) & = & \frac{1}{6}\left(C^{-1} \gamma_5 
\right)_ {\alpha_2 \alpha_3} \varepsilon_{\alpha_3 \alpha_4 \alpha_5 \alpha_6} 
\psi^{(3)}_{\alpha_4 \alpha_5 \alpha_6}(x) = 
\left\{\begin{array}{rll} 
\psi^{(3)}_{ 1 3 4}(x) & {\rm for} & \alpha_2 = 
1 \\ - \psi^{(3)}_{2 3 4}(x) & {\rm for} & \alpha_2 = 2 \\ 
\psi^{(3)}_{3 1 2}(x) & {\rm for} & \alpha_2 = 3 
\\ - \psi^{(3)}_{4 1 2}(x) & {\rm for} & \alpha_2 = 4  \end{array} \right. \; ,
\end{eqnarray} 

\ni   respectively,   where  $  \psi^{(N)}_{\alpha_1  \alpha_2  \ldots
\alpha_N}(x) $ for  active fermions  [Eq. (18)]  carries also  the \SM
(composite) label,  suppressed in  our  notation, and  $C$ denotes the
usual $4\times  4$ charge--conjugation matrix. We  can see that due to
the full antisymmetry in $\alpha_i $ indices for $i \geq 2$ these wave
functions or fields appear (up to the sign) with the multiplicities 1,
4, 24 and 1,  6 , respectively.  Thus, for active fermions and sterile
neutrinos there is given the weighting matrix 

\begin{equation} 
\rho^{(a)\,1/2} = \frac{1}{\sqrt{29}}  
\left( \begin{array}{ccc}  1  & 0 & 0 \\ 0 & 
\sqrt4 & 0  \\ 0 & 0 & \sqrt{24} \end{array} \right) 
\end{equation} 

\ni and

\begin{equation} 
\rho^{(s)\,1/2} = \frac{1}{\sqrt{7}}  
\left( \begin{array}{cc}  1  & 0  \\ 0 & \sqrt6 \end{array} \right) \; ,
\end{equation} 

\ni respectively. For all neutrinos ({\it i.e.}, $\nu_e\, ,\,\nu_\mu\,
,\,\nu_\tau $ and $ \nu_s\, ,\,\nu'_s$) described jointly, the overall
weighting matrix takes the form 

\begin{equation} 
\rho^{(a+s)\,1/2} = \frac{1}{\sqrt{36}}  
\left( \begin{array}{ccccc} 1 & 0 & 0 & 0 & 0 \\
0 & \sqrt6 & 0 & 0 & 0 \\ 0 & 0 & 1 & 0 & 0 \\  
0 & 0 & 0 & \sqrt4 & 0 \\ 0& 0 & 0 & 0 & \sqrt{24} \end{array} \right) \; ,
\end{equation} 

\ni if we use the ordering $s\, ,\,s'\, , \,e\, ,\,\mu\, ,\,\tau $. Of 
course, for all these matrices Tr $\rho = 1$.

 Concluding this  Introduction, we   would  like to  say that   in our
 approach to families of  fundamental particles Dirac bispinor indices
 ("algebraic  partons") play the role of  building blocks of composite
 states identified as fundamental particles. Any fundamental particle,
 {\it active}  with respect  to  the \SM  gauge interactions, contains
 {\it one} "active algebraic parton" (coupled to the \SM gauge bosons)
 and {\it  a number} $N-1$  of "sterile algebraic  partons" (decoupled
 from these bosons).  Due to the intrinsic  Pauli principle  obeyed by
 "sterile   algebraic partons",  the    number $N$  of all  "algebraic
 partons" within a fundamental particle is restricted by the condition
 $N-1 \leq 4$, so that only $N = 1,2,3,4,5$  are allowed. It turns out
 that states with $N = 1,3,5$ carry  total spin 1/2 and are identified
 with  three families of  leptons and  quarks, while states  with $N =
 2,4$ get total spin 0 and so far  are not identified. Any fundamental
 particle,  {\it sterile} with respect to  the \SM gauge interactions,
 contains  only a number $N  \leq  4$ of  "sterile algebraic partons",
 thus only $N = 1,2,3,4$ are allowed. States with $N = 1,3$ correspond
 to total spin 1/2 and have to be identified as two sterile neutrinos,
 while states with  $N = 2,4$ have total  spin 0 and  are still  to be
 identified. 

 Our algebraic construction   may   be interpreted {\it  either}    as
 ingeneously  algebraic   (much like  the   famous Dirac's   algebraic
 discovery of spin 1/2) {\it or} as the summit of an iceberg of really
 composite states of  $N$ spatial  partons with  spin  1/2 whose Dirac
 bispinor indices  manifest themselves as  our "algebraic partons". In
 the   former algebraic option,   we  avoid automatically  the irksome
 existence   problem of new   interactions  necessary to bind  spatial
 partons within  leptons and quarks  of the second and third families.
 For the latter spatial option  see  some remarks  in the second  Ref.
 [9]. 

\vspace{0.3cm}

\ni {\bf 2. A model of five--neutrino texture}

\vspace{0.3cm}

 In  this Section  we construct the  five--neutrino  mass matrix $ M =
 \left(M_{\alpha  \beta}   \right)  $ under  the   conjecture  that in
 ordering $\alpha\, ,\,\beta = s\, , \,s'\, , \,e\, , \,\mu\, , \,\tau
 $  of   neutrino sequence $\nu_\alpha    =  \nu_s\, ,  \,\nu'_s \,  ,
 \,\nu_e\,  , \,\nu_\mu\, ,  \,\nu_\tau $ only  the nearest neighbours
 are coupled through the matrix  $M$. In terms  of our presentation of
 three families  of   active fermions,  Eqs.  (18), and   two  sterile
 neutrinos,  Eqs. (19), such   an ordering of  five--neutrino sequence
 tells us that the chain 

$$
\nu_s \leftrightarrow \nu'_s \leftrightarrow \nu_e 
\leftrightarrow \nu_\mu \leftrightarrow \nu_\tau
$$

\ni of neutrino transitions corresponds to the chain

$$
\alpha_2 \leftrightarrow \alpha_2 \alpha_3 \alpha_4  
\leftrightarrow \alpha_1  \leftrightarrow 
\alpha_1 \alpha_3 \alpha_4  \leftrightarrow 
\alpha_1 \alpha_2 \alpha_3 \alpha_4 \alpha_5
$$

\ni  of consecutive acts  of creation  or  annihilation of index pairs
$\alpha_i \alpha_j  $ with $i,j  \geq 2$ (pairs of  "sterile algebraic
partons"),  allowed   by the   intrinsic  Pauli  principle   valid for
$\alpha_i $ with $i \geq  2$. In  one  of these four acts,  $\alpha_1$
must   additionally be  interchanged   with $\alpha_2  $,  what should
diminish   the rate of   $\alpha_1  \leftrightarrow  \alpha_2 \alpha_3
\alpha_4 $ {\it  versus}  $\alpha_1 \leftrightarrow  \alpha_1 \alpha_3
\alpha_4 $ ({\it i.e.}, the magnitude of $M_{s' e}$ {\it versus} $M_{e
\mu}$). One may also argue that  the rate of $\alpha_2 \leftrightarrow
\alpha_2  \alpha_3 \alpha_4 $  should  be  still more diminished  {\it
versus} $\alpha_1 \leftrightarrow  \alpha_1 \alpha_3 \alpha_4  $,({\it
i.e.}, $M_{s s'} $ {\it  versus} $M_{e \mu}$), as  being caused by two
such   additional interchanges of   $\alpha_1$ with  $\alpha_2 $.  The
allowed alternative chain 

$$
\alpha_1 \leftrightarrow \alpha_2 \alpha_3 \alpha_4  
\leftrightarrow \alpha_2  \leftrightarrow 
\alpha_1 \alpha_3 \alpha_4  \leftrightarrow 
\alpha_1 \alpha_2 \alpha_3 \alpha_4 \alpha_5
$$

\ni corresponding to

$$
\nu_e \leftrightarrow \nu'_s \leftrightarrow 
\nu_s \leftrightarrow \nu_\mu \leftrightarrow \nu_\tau
$$

\ni does not contain the natural link $\alpha_1  
\leftrightarrow \alpha_1 \alpha_2 \alpha_3 $ related to 
$\nu_e \leftrightarrow \nu_\mu $.

Thus, under the extra assumption that $ M_{s s} =0\, ,\,M_{s' s'} = 0$ 
and $M_{e e} = 0$, we can write

\begin{equation} 
M = \left( \begin{array}{ccccc}  0 & M_{s s'} & 0 & 0 & 0 \\  
M_{s' s} & 0  & M_{s' e} & 0 & 0 \\  0 & M_{e s'} & 0 & 
M_{e \mu} & 0 \\  0 & 0 & M_{\mu e} & M_{\mu \mu} &  
M_{\mu \tau} \\ 0& 0 & 0 & M_{\tau \mu} & M_{\tau \tau} 
\end{array} \right) \; ,
\end{equation} 

\ni where $M_{\beta \alpha} = M^*_{\alpha \beta}$ due to the hermicity
of $M$. When the CP violation may be ignored in neutrino oscillations,
we put $M^*_{\alpha \beta} =  M_{\alpha \beta}$ (and $M_{\alpha \beta}
> 0$ ). 

 Operating with the   mass matrix (23), we   will make the   tentative
 assumption   that,  in  comparison  with   its nonzero  off--diagonal
 entries, its nonzero diagonal elements are small enough to be treated
 as  a perturbation of  the former.   Such  a property  of $M$  may be
 related to   a tiny  neutrino mass scale   involved  in its  diagonal
 elements. Then, in the zero perturbative order (where $M_{\mu \mu} $,
 $ M_{\tau \tau}$  are put zero),  the matrix (23) can be diagonalized
 exactly,  giving     the   following zero--order     neutrino  masses
 ${\stackrel{o}{m}}_i$ numerated by $i = 4,5,1,2,3$: 

\begin{equation} 
\mi_4 = 0\, , \, \mi_{5,1} = \mp \left(A - \sqrt{B^2} 
\right)^{1/2}\, , \,\mi_{2,3} = \mp \left(A + \sqrt{B^2} \right)^{1/2}\, , 
\end{equation} 

\ni where

\vspace{-0.3cm} 

\begin{eqnarray}
2 A & = & |M_{e \mu} |^2 + |M_{\mu \tau} |^2  + |M_{s s'} |^2  + 
|M_{s' e} |^2 \; ,
\nonumber \\ 4 B^2 & = & \left( |M_{e \mu} |^2 + |M_{\mu \tau} |^2 - 
|M_{s s'} |^2  - |M_{s' e} |^2  \right)^2 + 4 |M_{\mu \tau} |^2 
|M_{s' e} |^2  \; . 
\end{eqnarray} 


\ni Next, in the first perturbative order with respect to the ratios

\begin{equation} 
\xi  \equiv M_{\tau \tau}/|M_{e \mu}|\; ,\; \chi 
\equiv M_{\mu \mu}/|M_{e \mu}|
\end{equation} 

\ni we obtain $ m_i = \mi_i + \delta m_i$, where

\begin{equation} 
\delta m_i = (C_i/D_i) |M_{e \mu}|
\end{equation} 

\ni with

\begin{eqnarray}
C_i & = & (\xi + \chi) \mi_i^4 -  \left[ \xi 
\left( |M_{e \mu} |^2 \!\! +\!\!  |M_{s s'} |^2 \!\!  + \!\!  
|M_{s' e} |^2  \right) + \chi \left( |M_{s s'} |^2  \!\!  + \!\!  
|M_{s' e} |^2  \right) \right] \mi_i^2  \nonumber \\ & & + \,\xi 
|M_{e \mu} |^2 |M_{s s'} |^2
\,, \nonumber \\ 
D_i & = & 5 \mi_i^4 - 3\left(|M_{e \mu} |^2 + |M_{\mu \tau} |^2  + 
|M_{s s'} |^2  + |M_{s' e} |^2  \right)\mi_i^2 + |M_{\mu \tau} |^2 
|M_{s' e} |^2 \,. 
\end{eqnarray}

\ni Note that the minus sign possible at $m_5$ and certain at $m_2$ is
irrelevant since $\nu_5 $ and $\nu_2 $  are relativistic particles for
which only $m^2_5$ and $m^2_2$ have physical meaning. 

 If our  argument outlined  in  the first   paragraph of this  Section
 works,  the  mass matrix elements  $M_{s s'}$  and $M_{s'  e}$ (which
 couple  the sterile neutrinos $\nu_s\,   , \,\nu'_s$ among themselves
 and $\nu'_s$ with the active $\nu_e$, respectively) should be smaller
 than the elements $M_{e \mu}$ and $M_{\mu \tau}$ (coupling the active
 neutrinos $\nu_e\,,\,\nu_\mu\,,\,\nu_\tau  $),  and also  the element
 $M_{s s'}$  should be smaller than $M_{s'  e}$: $|M_{s s'}|  < |M_{s'
 e}| < |M_{e  \mu}|$. Assuming tentatively $|M_{s  s'}| \ll |M_{e \mu}
 |$  and  $|M_{s' e}| \ll  |M_{e  \mu}|$, it can  be seen  that in the
 lowest approximation in the ratios 

\begin{equation} 
\lambda  \equiv |M_{s' e}|/|M_{e \mu}|\; ,\; \kappa 
\equiv |M_{s s'}|/|M_{e \mu}|
\end{equation} 

\ni the formulae (24) and (27) give

\begin{equation} 
\mi_4 = 0\, , \, \mi_{5,1} = 
\mp \left( \lambda^2 c^2 +  \kappa^2 \right)^{1/2} 
|M_{e \mu}|\, , \,\mi_{2,3} = \mp \frac{1}{s} |M_{e \mu}| 
\end{equation} 

\ni and

\begin{equation} 
\delta m_4 = \xi \frac{\kappa^2 s^2}{\lambda^2 c^2 + \kappa^2} 
|M_{e \mu}|\, , \, \delta m_{5,1} = 
\frac{1}{2}\xi \frac{\lambda^2 c^2 s^2}{\lambda^2 c^2 + \kappa^2} 
|M_{e \mu}|\, , \,\delta m_{2,3} = \frac{1}{2}(\xi c^2 + \chi) 
|M_{e \mu}| \, , 
\end{equation} 

\ni respectively, where the abbreviations

\begin{equation} 
s \equiv \frac{|M_{e \mu}| }{\left( |M_{e \mu}|^2 + 
|M_{\mu \tau}|^2 \right)^{1/2}}\, , \,
c \equiv \frac{|M_{\mu \tau}| }{\left( |M_{e \mu}|^2 + 
|M_{\mu \tau}|^2 \right)^{1/2}} =
(1 - s^2)^{1/2}
\end{equation} 

\ni are used.  Note that $\sum_i \delta   m_i = M_{\mu \mu}  + M_{\tau
\tau}$, as it should be because of $\sum_i \mi_i = 0$ and  $ M_{s s} =
M_{s' s'} = M_{e e} = 0 $. For the masses  $m_i = \mi_i + \delta m_i$,
the formulae (30) and (31)  show that $m^2_5 \stackrel{<}{\sim}  m^2_1
\ll m^2_2 \stackrel{<}{\sim} m^2_3$ and $m^2_4 \ll m^2_2$. 

 Now,  we  can calculate the  unitary  matrix $U =  \left(U_{\alpha i}
 \right)$   diagonalizing  the    mass  matrix  $M =   \left(M_{\alpha
 \beta}\right)$ given in Eq. (23): $U^\dagger  M U = {\rm diag} (m_4\,
 , \,m_5\,  , \,m_1\,  , \,m_2\, ,  \,m_3)$. In  the zero perturbative
 order  with  respect   to $\xi\,  ,\,    \chi  $ and  in the   lowest
 approximation in $\lambda\, , \,\kappa $, the result is 

\begin{equation} 
U = \left(   \begin{array}{ccccc}   f  & -   \frac{\kappa}{\lambda   c
\sqrt{2}}  f & \frac{ \kappa}{\lambda  c  \sqrt{2}} f & 0  &  0 \\ 0 &
\frac{1}{\sqrt{2}} & \frac{1}{\sqrt{2}} & \frac{\lambda s^2}{\sqrt{2}}
&  \frac{\lambda  s^2}{\sqrt{2}} \\  -  \frac{\kappa}{  \lambda} f & -
\frac{c}{\sqrt{2}} f &  \frac{c}{\sqrt{2}} f & -  \frac{s}{\sqrt{2}} &
\frac{s}{\sqrt{2}} \\   0   & - \frac{\lambda  s^2    }{\sqrt{2}}  & -
\frac{\lambda ^2}{ \sqrt{2}} & \frac{1}{\sqrt{2}} & \frac{1}{\sqrt{2}}
\\  \frac{\kappa    s}{\lambda  c}   f &   \frac{s}{\sqrt{2}}  f  &  -
\frac{s}{\sqrt{2}}   f &   -  \frac{c}{\sqrt{2}} &  \frac{c}{\sqrt{2}}
\end{array} \right) \; , 
\end{equation} 

\ni where

\begin{equation} 
f = \left(1 +  \frac{\kappa^2}{\lambda^2 c^2} \right)^{-1/2}\, , \,
\frac{\kappa}{\lambda c} f = \kappa \left(\lambda^2 c^2 + 
\kappa^2 \right)^{-1/2}
\end{equation} 

\ni assuming that $\lambda  c \neq 0 $. In  Eq. (33), a possible small
effect of  CP violation in neutrino oscillations  is ignored by taking
$M_{\alpha \beta} = |M_{\alpha \beta}|$. 

 If not only $M_{s s'} \ll M_{e \mu}$ and $M_{s' e} \ll M_{e \mu}$ but
 tentatively also $M_{s s'} \ll M_{s'  e}$, then beside $\kappa \ll 1$
 and $\lambda \ll  1$ also $\kappa \ll \lambda  $, and so,  we can put
 $\kappa = 0$ and $f = 1$. As is seen from Eq.  (33), in this case the
 sterile neutrino $\nu_s$ is decoupled from $\nu_s'\, , \,\nu_e\, , \,
 \nu_\mu\, , \, \nu_\tau  $ and, therefore, our five--neutrino texture
 is effectively reduced to a  four--neutrino texture, where the masses
 $m_i = \mi_i + \delta m_i$ given in Eqs. (30) and (31) become 

\begin{equation} 
m_4 = 0\, , \, m_{5,1} = \left( \mp \lambda c + \frac{1}{2} 
\xi s^2 \right) M_{e  \mu} \, , \,m_{2,3} = \left[ \mp \frac{1}{s} 
+ \frac{1}{2}\left( \xi c^2 + \chi \right) \right] M_{e \mu} \; .
\end{equation} 

\ni Here, $m^2_4 \stackrel{<}{\sim} m^2_5 \stackrel{<}{\sim} m^2_1 
\ll m^2_2 \stackrel{<}{\sim} m^2_3$.

 When the  effect   of mixing charged  leptons  $e^-\,  , \,\mu^-\,  ,
 \,\tau^-$ does  not    appear or may   be  ignored  in   the original
 lagrangian, then $V = U^\dagger $  is the five--neutrino extension of
 the  lepton  counterpart of  the familiar Cabibbo--Kobayashi--Maskawa
 mixing  matrix for quarks. In such  a situation, the flavor neutrinos
 $\nu_\alpha   $   and their  states $|\nu_\alpha    \rangle  $ can be
 expressed as 

\begin{equation} 
\nu_\alpha = \sum_i U_{\alpha i} \nu_i \, , \,|\nu_\alpha 
\rangle = \nu_\alpha^\dagger |0 \rangle = \sum_i 
U^*_{\alpha i} |\nu_i \rangle \, , 
\end{equation} 

\ni where  $\nu_i $  and $|\nu_i \rangle  $ are  massive neutrinos and
their states,  numerated by $  i  = 4,5,1,2,3$. Then,   in the case of
$\kappa^2 \ll \lambda^2  c^2$, for instance,  we obtain from Eqs. (33)
the following simple mixing of massive neutrinos: 

\begin{eqnarray}
\nu_s   &   = &   \nu_4  -   \frac{\kappa}{\lambda  c}  \frac{\nu_5  -
\nu_1}{\sqrt2}\;   ,  \nonumber  \\   \nu_s'   & =  &   \frac{\nu_5  +
\nu_1}{\sqrt2} + \lambda   s^2 \frac{\nu_2   +  \nu_3}{\sqrt2} \;    ,
\nonumber  \\ \nu_e  & = &  -c \left(  \frac{\nu_5 -  \nu_1}{\sqrt2} +
\frac{\kappa}{\lambda   c}  \nu_4    \right)   -  s    \frac{\nu_2   -
\nu_3}{\sqrt2}\;    ,  \nonumber  \\ \nu_\mu   &  =  &  \frac{\nu_2  +
\nu_3}{\sqrt2}  -  \lambda   s^2  \frac{\nu_5   + \nu_1}{\sqrt2}\;   ,
\nonumber \\ \nu_\tau  & = & s  \left( \frac{\nu_5 - \nu_1}{\sqrt2}  +
\frac{\kappa}{\lambda   c}   \nu_4 \right)    -    c  \frac{\nu_2    -
\nu_3}{\sqrt2}\; . 
\end{eqnarray}

\ni   Here,  we  have $f   =   1$ up to  $O(\kappa^2/\lambda^2  c^2)$.
Obviously, the assumption   $\kappa^2  \ll \lambda^2  c^2$  leading to
$\kappa^2 = 0$ is weaker than $\kappa  \ll \lambda c$ implying $\kappa
=  0$: in the   former case  the  sterile neutrino  $\nu_s $  is still
coupled to $\nu'_s\,  ,\,\nu_e\, ,\,\nu_\mu \, ,\,\nu_\tau $, although
by a small coefficient $ = O(\kappa/\lambda c)$. 

\vspace{0.3cm}

\ni {\bf 3. Five--neutrino oscillations}

\vspace{0.3cm}

 Finally, in this Section we can evaluate five--neutrino oscillation 
probabilities making use of the formulae

\begin{equation} 
P(\nu_\alpha \rightarrow \nu_\beta) = |\langle \nu_\beta| e^{iPL}|
\nu_\alpha\rangle|^2 = \delta_{\alpha \beta} - 4 \sum_{j>i} U^*_{ \beta
j} U_{\alpha j}U_{\beta i} U^*_{ \alpha i} \sin^2 x_{j i} \; ,
\end{equation} 

\ni valid when the CP violation may be ignored (then $U^*_{\alpha i} = 
U_{\alpha i}$). Here,

\begin{equation} 
x_{j i} = 1.27 \frac{\Delta m^2_{j i} L}{E}\; , \; \Delta m^2_{j i} = 
m^2_j - m^2_i
\end{equation}

\ni with $\Delta  m^2_{j i}$, $L$ and  $E$ expressed in eV$^2$, km and
GeV, respectively,   while  $p_i =  \sqrt{E^2   -  m^2_i}  \simeq  E -
m^2_i/2E$ are eigenvalues of  neutrino  momentum operator $P$ and  $L$
denotes the experimental baseline. 

 In the case of our mixing matrix (33), valid in the zero perturbative
 order  with  respect to  $\xi\,  ,  \,   \chi  $ and  in the   lowest
 approximation in $\lambda\, ,\, \kappa  $,  the formulae (38) in  the
 case of $\kappa^2 \ll \lambda^2 c^2$ give 

\begin{eqnarray}
P(\nu_e \rightarrow \nu_e)   & \!\!=\!\! &  1  - c^4 \sin^2 x_{1  5} -
(sc)^2\left( \sin^2 x_{2  1} \!+\! \sin^2 x_{3  1} \!+\!\sin^2 x_{2 5}
\!+\!\sin^2 x_{3  5}\right)  - s^4  \sin^2   x_{3 2}   \nonumber \\  &
\!\simeq\! &  1 - c^4\sin^2 x_{1  5} - (2s  c)^2 \sin^2 x_{2 1}  - s^4
\sin^2 x_{3 2}  \;  , \nonumber \\  P( \nu_\mu  \rightarrow \nu_\mu) &
\!\!=\!\! & 1 - \sin^2 x_{3 2} \; , \nonumber \\ P(\nu_\mu \rightarrow
\nu_e) & \!\!=\!\! & s^2 \sin^2 x_{3 2}  - \lambda c s^3 \left( \sin^2
x_{2 1}  - \sin^2 x_{3  1} - \sin^2 x_{2 5}  + \sin^2  x_{3 5} \right)
\nonumber \\ & \!\!\simeq\!\! & s^2 \sin^2 x_{3 2}\, . 
\end{eqnarray}

\ni Here, due to Eqs. (38) and (34) we have $\Delta m^2_{2 1} 
\simeq \Delta m^2_{3 1} \simeq \Delta m^2_{2 5} \simeq \Delta m^2_{3 5}$ and 

\begin{equation} 
\Delta m^2_{1 5} = 2 \xi \lambda s^2 c M^2_{e \mu}\; , \; 
\Delta m^2_{3 2} =2\frac{\xi c^2 + \chi}{s} M^2_{e \mu}\; , \; 
\Delta m^2_{2 1} = \frac{1}{s^2} M^2_{e \mu}\; .
\end{equation}

 When $1.27   \Delta m^2_{3 2}  L_{\rm  atm}/E_{\rm  atm}=  O(1)$  for
 atmospheric $\nu_\mu$'s and thus  $\Delta m^2_{3 2} = \Delta m^2_{\rm
 atm} \sim 3.5  \times  10^{-3}\;{\rm eV}^2$  [4], the second  formula
 (40) is able to describe atmospheric neutrino oscillations (dominated
 in our case by  the mode $  \nu_\mu \leftrightarrow \nu_\tau  $) with
 maximal  amplitude 1. Thus, the  second   equation (40) leads to  the
 estimation 

\begin{equation} 
2\frac{\xi c^2 + \chi}{s} M^2_{e \mu} \sim 3.5 \times 10^{-3} {\rm eV}^2 \; .
\end{equation}

\ni Hence, $\xi + \chi \rightarrow 0$ with $c \rightarrow 1$ for $M_{e
\mu}$ fixed. Also  $ \xi $ and $\chi  \rightarrow 0$, since $\xi $ and
$\chi \geq 0$. 

 On the other hand, when  $\Delta m^2_{1 5}  L_{\rm sol}/E_{\rm sol} =
 O(1)$ for solar $\nu_e$'s and so, $\Delta m^2_{1 5} = \Delta m^2_{\rm
 sol} \sim  (6.5\times 10^{-11}$ or $4.4\times  10^{-10})\;{\rm eV}^2$
 [3] (when  considering the "small" or  "large" vacuum solution), then
 the first   formula (40)  has   a chance to describe   solar neutrino
 oscillations  (dominated now by  the mode $\nu_e \rightarrow \nu'_s$)
 with the large amplitude  $c^4 = \sin^2  2\theta_{\rm sol} \sim 0.72$
 or 0.90, respectively. In fact, due to  $ \Delta m^2_{1 5} \ll \Delta
 m^2_{32} \ll \Delta m^2_{2 1}$ the first formula (40) becomes 

\begin{equation} 
P(\nu_e \rightarrow \nu_e) \simeq 1 - c^4  \sin^2 x_{1 5} - 
\left( 2s^2 c^2 + s^4/2 \right) \; ,
\end{equation}

\ni where the  disturbing last term, $2s^2  c^2 + s^4/2  \sim 0.27$ or
0.099, may  be too large,   but it tends quickly   to zero  with  $c^4
\rightarrow 1$. Thus, from the first equation (41) the estimate 

\begin{equation} 
2 \xi \lambda s^2c M^2_{e \mu} \sim  \left( 6.5 \times 10^{-11}\;\,
{\rm or}\;\,4.4 \times  10^{-10} \right)\,{\rm eV}^2
\end{equation}

\ni is suggested.

 Since $ c^2 \sim 0.85$ or 0.95 and $ c \sim 0.92 $ or 0.97, Eqs. (42)
 and (44) in the case of $ \xi \gg \chi $ ({\it i.e.}, $ M_{\tau \tau}
 \gg M_{\mu \mu} $ ) give the estimation 

\begin{equation} 
\xi M^2_{e \mu} \sim  \left(8.0 \;\,{\rm or}\;\, 4.2 \right) 
\times 10^{-4}\, {\rm eV}^2 \; , \; \lambda \sim  2.9\times 
10^{-7} \;\,{\rm or}\;\, 1.1\times 10^{-5}\, .
\end{equation}

\ni  Such  a tiny value  $\lambda  $ shows  that $M_{s'  e}$ (coupling
$\nu'_s$ with  $\nu_e$) is    really very  small  {\it  versus}  $M_{e
\mu}:\;\,M_{s' e} = \lambda M_{e \mu}$. In this case, we get from Eqs.
(35) 

\vspace{-0.3cm}

\begin{eqnarray} 
m_{5,1}: m_{2,3}\! \simeq \!\lambda  sc  \mp \frac{1}{2} \xi s^3  \sim
\left(1.0 \times  10^{-7}\!\!\right. &  {\rm or} & \!\!\left.2.3\times
10^{-6} \right) \mp 2.4 \times  10^{-6} \, ({\rm eV}/M_{e \mu})^2\,  ,
\nonumber \\ m_{2,3} \!=\! \mp \frac{1}{s} M_{e  \mu}\! & \!\sim & \mp
(2.6\;{\rm or}\; 4.4) M_{e \mu} \; . 
\end{eqnarray} 

\ni Thus, in  order  to obtain, for instance,  $|m_2|  \simeq m_3 \sim
(1\;{\rm to}\;10)  $~eV one  should  take $M_{e \mu}  \sim (0.38\;{\rm
to}\; 3.8)$ or $(0.23\;{\rm to}\;  2.3)$ eV. Then,  from the first Eq.
(44) we infer that 

\begin{equation} 
\xi  \sim 5.6 \times \left(10^{-3}\; {\rm to} \;10^{-5}\right)\; 
{\rm or}\; 7.9\times \left( 10^{-3}\; {\rm to} \;10^{-5}\right) 
\end{equation}

\ni for $c^4 \sim 0.72$ or 0.90, respectively. However, when $M_{e \mu}$ 
is kept fixed, $\xi $ tends quickly to zero with $c^4 \rightarrow 1$.

 In the case of Chooz experiment searching for oscillations of reactor
 $\bar{\nu}_e $'s [13], where  it  happens that $1.27 \Delta  m^2_{\rm
 atm} L_{\rm Chooz}/E_{\rm Chooz} = O(1)$, the first formula (40) with
 $\Delta m^2_{3 2} = \Delta  m^2_{\rm atm} $ and  $\Delta m^2_{1 5}  =
 \Delta m^2_{\rm sol}$ becomes 

\begin{equation} 
P(\bar{\nu}_e \rightarrow \bar{\nu}_e) 
\simeq 1 - s^4  \sin^2 x_{3 2} - 2s^2 c^2  
\simeq 1 - \left( 2s^2 c^2 + s^4 \right) \; ,
\end{equation}

\ni  since $\Delta  m^2_{1 5} \ll  \Delta  m^2_{3 2} \ll \Delta m^2_{2
1}$.  This is  consistent   with  the negative  result  $P(\bar{\nu}_e
\rightarrow \bar{\nu}_e) = 1$ of Chooz experiment  up to 28\% or 10\%,
but this deviation from 1 tends quickly  to zero with $c^4 \rightarrow
1$. Note that $U_{e 3} = s/\sqrt{2} \sim 0.28$ or 0.16, respectively. 

 The third  formula   (40)   may  imply  the existence    of  $\nu_\mu
 \rightarrow \nu_e$ oscillations with the amplitude equal to $s^2 \sim
 0.15$ or 0.05 and the mass--square scale given by $\Delta m^2_{3 2} =
 \Delta m^2_{\rm atm} \sim 3.5  \times 10^{-3}\;{\rm eV}^2$, while the
 estimate from LSND experiment [2] is, say, $\sin^2 2\theta_{\rm LSND}
 \sim 0.02$ and $\Delta m^2_{\rm  LSND}  \sim 0.5\;{\rm eV}^2$.  Thus,
 our four--neutrino texture,    if  fitted to atmospheric  and   solar
 results, cannot explain the LSND observation. In order to include the
 LSND   effect,     one  might   depart   from   our   conjecture   on
 nearest--neighbour   coupling  in the  four--  or five--neutrino mass
 matrix. 

\vspace{0.3cm}

\ni {\bf 4. Final remarks: a specific proposal for mass matrix elements}

\vspace{0.3cm}

 In a specific model of three--neutrino texture discussed by the author 
previously ({\it e.g.} [5] and the first Ref. [6]), the following nonzero 
mass matrix elements were proposed:

\vspace{-0.1cm}

\begin{equation} 
M_{\mu \mu} = \frac{4\cdot 80}{9} \frac{\mu}{29}\, ,\,M_{\tau \tau} = 
\frac{24\cdot 624}{25} \frac{\mu}{29}\, ,\,M_{e \mu} = 
\frac{2 g}{29}\, ,\,M_{\mu \tau} = \frac{8\sqrt {3}\, g}{29}\, ,
\end{equation}

\ni where $\mu $ and $g$ stood for two small mass scales. Then,

\vspace{-0.1cm}

\begin{equation} 
M_{\tau \tau} = 16.848\, M_{\mu \mu}\, ,\,M_{\mu \tau} = 
\sqrt {48}\, M_{e \mu}
\end{equation}

\ni and from Eqs. (26)

\vspace{-0.1cm}

\begin{equation} 
\xi = 299.52\,\mu /g \, ,\, \chi = 17.778\, \mu /g = \xi /16.848\, .
\end{equation}

\ni Thus, $\xi \ll 1$ if and only if $ g \gg 299.52  \mu $, the latter
inequality  implying $g \gg \mu $  certainly. In the zero perturbative
order with respect to $\xi $ or $\mu /g$ we put $ \mu = 0$. 

 When accepting the values (49) for $M_{e \mu}$ and $M_{\mu \tau}$, we 
obtain from Eqs. (32)

\begin{equation} 
s = 1/7 = 0.14286 \, ,\, c = \sqrt {48}/7 = 0.98974 \, .
\end{equation}

\ni So, the estimation (42) provided by atmospheric neutrino experiments 
implies

\begin{equation} 
\xi M^2_{e \mu} \sim 2.4 \times 10^{-4}\;{\rm eV}^2 \, .
\end{equation}

\ni Then, due to Eq. (43) and the first Eq.  (41), the mass--square 
difference and oscillation amplitude for solar neutrinos should be

\begin{equation} 
 \Delta m^2_{\rm sol} = 2 \xi \lambda s^2c M_{e \mu} \sim 9.7 
\times 10^{-6}\,\lambda\; {\rm eV}^2 \, ,\, \sin^2 \theta_{\rm sol} 
= c^4 = 0.95960 \, ,
\end{equation}

\ni respectively, while the disturbing last term would become smaller 
than before, giving now $2 s^2 c^2 + s^4/2 \sim 0.040$.

 The values  (49)  proposed for  elements of the  three--neutrino mass
 matrix $M^{(a)} =  \left( M_{\alpha \beta} \right)$ $(\alpha,\beta  =
 e\,   , \,\mu\, ,   \,\tau)$ can be   exactly deduced from the simple
 ansatz ([10] and the first Ref. [6]): 

\begin{equation} 
M^{(a)} = \rho^{(a) 1/2}\left[ \mu(N^2 - N^{-2}) + 
g(a + a^\dagger)\right]\rho^{(a) 1/2} \, .
\end{equation}

\ni Here,

\vspace{-0.1cm}

\begin{equation} 
N = 1 + 2 a^\dagger a = \left( \begin{array}{ccc} 1 & 0 & 0 \\ 0 & 3 & 0 \\ 0
& 0 & 5 \end{array}\right)  
\end{equation}

\ni is the matrix of number of all Dirac bispinor indices $\alpha_i$ (all 
"algebraic partons") used in Eqs. (18) to present active neutrinos 
$\nu_e\, , \,\nu_\mu\, , \,\nu_\tau $, while

\vspace{-0.1cm}

\begin{equation} 
a = \left( \begin{array}{ccc} 0 & 1 & 0 \\ 0 & 0 & \sqrt2 
\\ 0 & 0 & 0 \end{array}\right)  \; ,\;a^\dagger = 
\left( \begin{array}{ccc} 0 & 0 & 0 \\ 1 & 0 & 0 \\ 0 & 
\sqrt2 & 0 \end{array}\right)  
\end{equation}

\ni are (truncated) annihilation and creation  matrices of index pairs
$\alpha_i  \alpha_j $ with $ i,j  \geq 2$ (pairs of "sterile algebraic
partons") included in  Eqs.   (18) for  active neutrinos. The   latter
matrices  satisfy,  jointly with  the matrix  of  number of such index
pairs, 

\begin{equation} 
n = a^\dagger a = \left( \begin{array}{ccc} 0 & 0 & 0 
\\ 0 & 1 & 0 \\ 0 & 0 & 2 \end{array}\right) \, , 
\end{equation}

\ni the familiar commutation relations 

\vspace{-0.1cm}

\begin{equation} 
[a\, , \,n] = a \;,\; [a^\dagger\, , \,n] = -a^\dagger\, ,
\end{equation}

\ni  and,  in  addition,   the  truncation relations   $a^3  = 0$  and
$a^{\dagger 3} = 0$ consistent with  the intrinsic Pauli principle for
Dirac bispinor indices $\alpha_i $ with $i \geq 2$ (obviously, neither
boson   nor fermion    canonical  commutation   relations,   $[a\,   ,
\,a^\dagger]_\mp = 1$, are satisfied here). Finally, $\rho^{(a) 1/2} $
stands in Eq. (55) for the active--neutrino weighting matrix (20). 

 In the mass matrix  (55), the first  term containing $\mu N^2$ may be
 intuitively  interpreted as  an   interaction of all  $N$  "algebraic
 partons"  treated on equal  footing, while the second involving $-\mu
 N^{-2}$, as a subtraction term  caused by the  fact that there is one
 "active algebraic  parton" distinguished  (by its  external coupling)
 among all $N$  "algebraic partons" of  which $N-1$, as "sterile"  are
 undistinguishable. This  distinguished    "algebraic parton" appears,
 therefore, with   the probability $[N!/(N-1)!]^{-1}   = N^{-1}$ that,
 when squared,  leads  to an  additional   interaction involving  $\mu
 N^{-2}$. The latter interaction should be  subtracted from the former
 in order to obtain for  $N = 1$ the matrix  element $M_{e e}$ assumed
 to be zero. The third term in the mass matrix (55) containing $g (a +
 a^\dagger)$ annihilates   and  creates  pairs  of "sterile  algebraic
 partons" and so, is responsible in a natural way  for mixing of three
 active neutrinos. 

 Of course, the  three--neutrino matrix  $M^{(a)}  = \left(  M_{\alpha
 \beta} \right)$ $(\alpha,\beta = e\,  , \,\mu\, , \,\tau)$ considered
 in this Section is a submatrix of our five--neutrino mass matrix $M =
 \left(  M_{\alpha  \beta} \right)$ $(\alpha,\beta =   s\,  , \,s'\, ,
 \,e\, , \,\mu\, , \,\tau)$, {\it viz.} 

\vspace{-0.1cm}

\begin{equation} 
\!M \!=\! \left(\!  \begin{array}{ccccc} 0 &  M_{s s'} & 0  & 0 & 0 \\
M_{s' s} & 0 & 0 & 0 & 0 \\ 0 & 0 & 0 & 0 & 0 \\ 0 & 0 & 0  & 0 & 0 \\
0&  0    &  0  &   0    & 0  \end{array}\!  \right)\!     + \!\left(\!
\begin{array}{ccccc} 0 & 0 & 0 & 0 & 0 \\ 0 & 0 & M_{s' e}  & 0 & 0 \\
0 & M_{e s'} & 0 & 0 & 0  \\ 0 &  0 & 0 & 0  & 0 \\  0& 0 &  0 & 0 & 0
\end{array} \!\right)\! + \!\left(\! \begin{array}{ccccc}  0 & 0 & 0 &
0 & 0 \\ 0 & 0 & 0 &  0 & 0 \\ 0  & 0 & 0  & M_{e \mu} & 0  \\ 0 & 0 &
M_{\mu e} &  M_{\mu \mu} & M_{\mu  \tau} \\ 0& 0 &  0 & M_{\tau \mu} &
M_{\tau \tau} \end{array}\! \right) \,, 
\end{equation} 

\ni where $M_{s s'} = \kappa M_{e  \mu}$ and $M_{s'  e} = \lambda M_{e
\mu}$. Thus, the $2\times 2$ matrix involved in the middle $5\times 5$
matrix plays  the role of  coupling  between the sterile  $2\times  2$
matrix $M^{(s)}$ and active  $3\times 3$ matrix $M^{(a)}$. If $\lambda
$ were zero, both  sterile neutrinos  $\nu_s  $ and $\nu'_s$ would  be
decoupled from the three active. When $\kappa  $ is put zero, $\nu_s $
becomes decoupled from $\nu'_s$ as well as from $\nu_e\, , \,\nu_\mu\,
, \,\nu_\tau $ ($M^{(s)}$ is then a zero matrix). 

 Originally, the ansatz (55) was  introduced for mass matrix $ M^{(e)}
 =  \left(  M^{(e)}_{\alpha \beta}   \right)$ $(\alpha,\beta =   e\, ,
 \,\mu\, , \,\tau)$ of charged leptons $ e^-\,  , \,\mu^-\, , \,\tau^-
 $.   In this case,  in order  to get  a  small  but  nonzero value of
 $M^{(e)}_{e  e}$, the  quantity  $-\mu(1- \varepsilon)N^{-2}$  with a
 small  $\varepsilon$ (rather than  the  quantity $-\mu N^{-2} $)  was
 used in the  second term of $M^{(e)} $.Then,  the nonzero mass matrix
 elements were 

\begin{eqnarray} 
M_{e e}^{(e)} = \varepsilon  \frac{\mu}{29}\, , \, M_{\mu \mu}^{(e)} &
=    & \frac{4 (80+\varepsilon)}{25}    \frac{\mu}{29} \, , \, M_{\tau
\tau}^{(e)} = \frac{24  (624  +  \varepsilon)}{25}\frac{\mu}{29} \,  ,
\nonumber  \\  M_{e \mu}^{(e)} & =  &  \frac{2  g}{29}  \, , \, M_{\mu
\tau}^{(e)} = \frac{8\sqrt{3} g}{29} \, . 
\end{eqnarray} 

\ni  Making the conjecture that  for charged leptons diagonal elements
of $  M^{(e)} $ dominate over its  off--diagonal entries  ({\it i.e.},
$\xi \gg 1$ what  is certainly true for $g  \ll \mu $),  we calculated
the masses $m_e\, , \,m_\mu\, , \,m_\tau$  as eigenvalues of $M^{(e)}$
in the lowest (quadratic) perturbative order  with respect to $1/\xi $
or $g/\mu $. Then, we expressed $m_\tau$, $\mu $ and $\varepsilon $ in
terms of $m_e\, , \,m_\mu $ and $(g/\mu)^2$, obtaining 

\vspace{-0.2cm}

\begin{eqnarray} 
m_\tau & = & \left[1776.80 + 10.12112(g/\mu)^2 \right] \; 
{\rm MeV} \; ,\nonumber \\ \mu & = & \left\{ 85.9924 + 
O\left[(g/\mu)^2\right] \right\} \; {\rm MeV} \; , \nonumber \\ 
\varepsilon & = & 0.172329 + O\left[(g/\mu)^2\right]  \; ,  
\end{eqnarray} 

\ni where the experimental values of $M_e$  and $M_\mu $ were taken as
the only  input.  Comparing this  prediction for  $m_\tau  $ with  the
experimental value  $m^{\rm exp}_\tau  =  1777.05^{+0.29}_{-0.26}$ MeV
[14], we got 

\begin{equation} 
(g/\mu)^2 = 0.024^{+0.028}_{-0.025}
\end{equation}

\ni for charged leptons.  In such a way, we   achieved in the case  of
charged leptons a  really good agreement of  our ansatz  for $M^{(e)}$
with the experimental mass   spectrum (even in the   zero perturbative
order). 

 This result has motivated the application of our ansatz for $M^{(e)}$
 also  to the  case   of active  neutrinos  $\nu_e\, ,   \,\nu_\mu\, ,
 \,\nu_\tau$ (corresponding  to $ e^-\, ,  \,\mu^-\, , \,\tau^- $). In
 their case, however, the inverse conjecture $\xi \ll 1$ or $g \gg \mu
 $ seems natural in  view  of experimentally suggested  large neutrino
 mixing that   is  in contradistinction to  small  mixing   of charged
 leptons  [{\it cf.} Eq.  (63)].  In terms of  three  active neutrinos
 alone this ansatz leads to maximal amplitude for atmospheric $\nu_\mu
 \rightarrow  \nu_\mu $  oscillations, but it  requires introducing at
 least one sterile neutrino to  explain solar $\nu_e \rightarrow \nu_e
 $ oscillations ([5]  and the present paper).  Even in this  case, the
 LSND  effect does   not appear,  however.   Thus, if this effect  was
 confirmed   in a clear   manner, the conjecture on nearest--neighbour
 coupling  in   the  four-- or  five--neutrino  mass   matrix  and, in
 particular, our ansatz (55) would not be correct for neutrinos. 

 In this  case, a  different neutrino texture,   also including one or
 more  sterile neutrinos,  would be needed.   If, on the contrary, the
 LSND effect was not seen, our  four-- or five--neutrino texture might
 be realized in Nature. However, much more  economical would be then a
 three-neutrino texture involving (as  {\it e.g.} in Ref. [15]) active
 neutrinos $\nu_e\,  ,   \,\nu_\mu\,  ,  \,\nu_\tau$   with the   mass
 hierarchy $ m^2_1 \stackrel{<}{\sim} m^2_2   \ll m^2_3$ (in place  of
 $m^2_1 \ll m^2_2    \stackrel{<}{\sim} m^2_3$ valid   in  the present
 paper). They ought  to  be coupled in a  different  way than in   the
 present  paper in  order to  explain both the  atmospheric and  solar
 neutrino results.  The argument for our  texture would be the absence
 of LSND effect and,  at the same time,  the experimental existence of
 one or two sterile neutrinos  (for a possible astrophysical aspect of
 sterile neutrinos {\it cf. e.g.} Ref. [8] ). 

\vfill\eject

~~~~
\vspace{0.5cm}

{\centerline{\bf References}}

\vspace{0.3cm}

{\everypar={\hangindent=0.5truecm}
\parindent=0pt\frenchspacing

{\everypar={\hangindent=0.5truecm}
\parindent=0pt\frenchspacing

~1.~{\it Cf. e.g.} C.~Giunti, hep--ph/990336; C.~Giunti, 
M.C.~Gonzalez--Garcia and
Pe\~{n}a--Garay, hep--ph/0001101; and references therein.

\vspace{0.06cm}

~2.~C.~Athanassopoulos {\it et al.} (LSND Collaboration), 
{\it Phys. Rev. Lett.} {\bf 75}, 2650 (1995); {\it Phys. Rev.} 
{\bf C 54}, 2685 (1996); {\it Phys. Rev. Lett.} {\bf 77}, 
3082 (1998); {\bf 81}, 1774 (1998).

\vspace{0.06cm}

~3.~{\it Cf. e.g.}~J.N.~Bahcall, P.I.~Krastev and A.Y.~Smirnov, 
{\it Phys. Rev.} {\bf D 58}, 096016 (1998); hep--ph/9905220v2; 
{\it Phys. Lett.} {\bf B 477}, 401 (2000); hep--ph/0002293.

\vspace{0.06cm}

~4.~Y.~Fukuda {\it et al.} (Super--Kamiokande Collaboration), 
{\it Phys. Rev. Lett.} {\bf 81}, 1562 (1998) [Errata {\bf 81}, 
4279 (1998)]; {\bf 82}, 1810 (1999);  {\bf 82}, 2430 (1999).

\vspace{0.06cm}

~5.~W. Kr\'{o}likowski, hep--ph/0001023.

\vspace{0.06cm}

~6.~{\it Cf. e.g.} W. Kr\'{o}likowski, {\it Nuovo Cim.} {\bf A 122},
893 (1999); {\it Acta Phys. Pol.} {\bf B 30}, 2631 (1999); {\bf B 31}, 
663 (2000); and references therein.

\vspace{0.06cm}

~7.~M.~Gell--Mann, P.~Ramond and R.~Slansky, in {\it Supergravity }, ed.
P.~van Nieuwenhuizen and D.Z.~Freedman, North--Holland, Amsterdam 1979;
T.~Yanagida, in {\it Proc. of the Workshop on the Unified Theory of the
Baryon Number in the Universe}, ed. O.~Sawada and S.~Sugamoto, KEK report
No. 79--18, Tsukuba (Japan) 1979; see also R.~Mohapatra and G.~Senjanovic,
{\it Phys. Rev. Lett.} {\bf 44}, 912 (1980).

\vspace{0.06cm}

~8.~D.O.~Caldwell, hep--ph/0003250.

\vspace{0.06cm}

~9.~W. Kr\'{o}likowski, {\it Acta Phys. Pol.} {\bf B 30}, 227 (1999); 
in {\it Proc. of the 12th Max Born Symposium, Wroc{\l}aw (Poland) 1998}, 
eds. A.~Borowiec  {\it et al.}, Springer 2000, also hep--ph/9808307.

\vspace{0.06cm}

10.~W. Kr\'{o}likowski, {\it Acta Phys. Pol.} {\bf B 21}, 871 (1990); 
{\it Phys. Rev.} {\bf D 45}, 3222 (1992); in {\it Proc. of the 2nd 
Max Born Symposium, Wroc{\l}aw (Poland) 1992}, eds. Z.~Oziewicz  
{\it et al.}, Kluwer 1993; {\it Acta Phys. Pol.} {\bf B 24}, 1149 (1993).

\vspace{0.06cm}

11.~T. Banks, Y. Dothan and D.~Horn, {\it Phys. Lett.} 
{\bf B 117}, 413 (1982).

\vspace{0.06cm}

12.~E. K\"{a}hler, {\it Rendiconti di Matematica} {\bf 21}, 425 (1962); 
see also D.~Ivanenko and L.~Landau, {\it Z. Phys.} {\bf 48}, 341 (1928).

\vspace{0.06cm}

13.~M. Appolonio {\it et al.} (Chooz Collaboration), {\it Phys. Lett.} 
{\bf B 420}, 397 (1998)

\vspace{0.06cm}

14.~Particle Data Group, {\it Review of Particle Physics, Eur. Phys.~J.} 
{\bf C 3}, 1 (1998).

\vspace{0.06cm}

15.~P.H. Chankowski, W. Kr\'{o}likowski and S.~Pokorski, {\it Phys. Lett.} 
{\bf B 473}, 109 (2000).

\vfill\eject

\end{document}